\documentclass[twocolumn]{aastex631}
\usepackage[T1]{fontenc}

\usepackage[version=4]{mhchem}

\usepackage{ulem}

\shorttitle{Primordial Origin of Methane on Eris and Makemake}
\shortauthors{Mousis et al.}
\graphicspath{{./}{figures/}}

\DeclareUnicodeCharacter{2212}{-}
\begin{document}

\title{Primordial Origin of Methane on Eris and Makemake Supported by D/H Ratios}

\correspondingauthor{Olivier Mousis}
\email{olivier.mousis@lam.fr}

\author[0000-0001-5323-6453]{Olivier Mousis}
\affiliation{Aix-Marseille Universit\'e, CNRS, CNES, Institut Origines, LAM, Marseille, France}
\affiliation{Institut Universitaire de France (IUF), France}
\author[0009-0005-1133-7586]{Aaron Werlen}
\affiliation{Aix-Marseille Universit\'e, CNRS, CNES, Institut Origines, LAM, Marseille, France}
\affiliation{Institute for Particle Physics and Astrophysics, ETH Zurich, Zurich, Switzerland}
\author[0000-0002-8719-7867]{Tom Benest Couzinou}
\affiliation{Aix-Marseille Universit\'e, CNRS, CNES, Institut Origines, LAM, Marseille, France}
\author[0000-0002-3289-2432]{Antoine Schneeberger}
\affiliation{Aix-Marseille Universit\'e, CNRS, CNES, Institut Origines, LAM, Marseille, France}



\begin{abstract}
Deuterium, a heavy isotope of hydrogen, is a key tracer of the formation of the Solar System. Recent JWST observations have expanded the dataset of D/H ratios in methane on the KBOs Eris and Makemake, providing new insights into their origins. This study examines the elevated D/H ratios in methane on these KBOs in the context of protosolar nebula dynamics and chemistry, and proposes a primordial origin for the methane, in contrast to previous hypotheses suggesting abiotic production by internal heating. A time-dependent disk model coupled with a deuterium chemistry module was used to simulate the isotopic exchange between methane and hydrogen. Observational constraints, including the D/H ratio measured in methane in comet 67P/Churyumov-Gerasimenko, were used to refine the primordial D/H abundance. The simulations show that the observed D/H ratios in methane on Eris and Makemake are consistent with a primordial origin. The results suggest that methane on these KBOs likely originates from the protosolar nebula, similar to cometary methane, and was sequestered in solid form --either as pure condensates or clathrates-- within their building blocks prior to accretion. These results provide a { simple} explanation for the high D/H ratios in methane on Eris and Makemake, without the need to invoke internal production mechanisms.
\end{abstract}



\section{Introduction} 
\label{sec:sec1}

Deuterium, a heavy isotope of hydrogen, was produced during the birth of the Universe with a primordial D/H abundance of $\sim$2.5 $\times$ 10$^{-5}$ \citep{Kisl24} and is subsequently destroyed in stellar interiors. As a key tracer of the formation and evolution of the Solar System, the D/H ratio varies with temperature and chemical conditions, providing critical insights into the thermodynamic state of the early protosolar nebula (PSN) and the origin of planetary bodies. In the cold outer regions of the PSN, deuterium fractionation produces deuterium--enriched ices and organics, typically close to values observed in the interstellar medium (ISM), while the warmer inner regions { acquire} lower D/H ratios. This creates a spatial and thermal record of the protoplanetary disk, preserved in the D/H ratios measured across different reservoirs of the Solar System \citep{Dr99,Mo00,Ce14}.

Recent observations with the James Webb Space Telescope (JWST) revealed the presence of monodeuterated methane (CH$_3$D) on the surfaces of the KBOs Eris and Makemake. The measured D/H ratios in methane were (2.5 $\pm$ 0.5) $\times$ 10$^{-4}$ for Eris and (2.9 $\pm$ 0.6) $\pm$ 10$^{-4}$ for Makemake \citep{Gr24}. Previously, D/H measurements in CH$_4$ were limited to Titan's atmosphere, where the Cassini Composite Infrared Spectrometer determined a value of (1.59 $\pm$ 0.33) $\times$ 10$^{-4}$ \citep{Ni12}, and a small number of comets. For comets C/2004 Q2 (Machholz), C/2001 Q4 (NEAT), and C/2007 N3 (Lulin), only upper limits were reported, with D/H values of $<$ 0.005, $<$ 0.01, and $<$ 0.0075, respectively \citep{Ka05,Bo09,Gi12}. In contrast, a precise D/H measurement of (2.41 $\pm$ 0.29) $\times$ 10$^{-3}$ was obtained for comet 67P/Churyumov–Gerasimenko (67P/C-G) using the ROSINA instrument aboard the Rosetta mission \citep{Mu22}.

The D/H measurements in CH$_4$ on Makemake and Eris, as interpreted by \cite{Gl24}, provide evidence for abiotic methane production. According to this interpretation, CH$_4$ production may have been triggered by significant radiogenic heating, potentially driving hydrothermal circulation at the base of an ice-covered ocean and facilitating geochemical transformations of volatile species. \cite{Gl24} also suggest that metamorphic reactions involving accreted organic matter may have occurred due to heating in the deeper interior, resulting in the production of thermogenic methane.

Here, we propose an alternative interpretation, suggesting that the D/H ratios measured in methane on Makemake and Eris align with a primordial origin, similar to the CH$_4$ observed in numerous comets. In this scenario, methane on these KBOs would have originated in the PSN and been accreted during their formation, similar to the process hypothesized for methane in comets. To investigate this hypothesis, we use a time-dependent protoplanetary disk model coupled with a module that simulates the deuterium exchange between CH$_4$ and H$_2$ in the gas phase of the PSN. A key uncertainty in this analysis is the primordial deuterium abundance acquired by CH$_4$ from the ISM during its incorporation into the PSN, since no reliable measurements of the D/H ratio in interstellar methane exist. To date, the only tentative ISM detection potentially associated with CH$_3$D does not precisely match any cataloged spectral lines \citep{Sa12}. Consequently, we adopt the D/H ratio measured in comet 67P/C-G as a conservative estimate for our calculations.

Section \ref{sec:sec2} provides a detailed description of the PSN model and its accompanying deuterium chemistry module. Section \ref{sec:sec3} presents the simulation results, while Section \ref{sec:sec4} outlines the conclusions and discusses their implications.

\section{Model} 
\label{sec:sec2}

\subsection{Model of Protosolar Nebula}
\label{sec:sec2.1}

To compute the thermodynamic evolution of the PSN, we use a one-dimensional $\alpha$-viscous accretion disk model based on the work of \cite{Ag20}. The viscosity of the disk is given by $\nu$ = $\alpha c_s^2/\Omega_K$, following the prescription of \cite{sh73}, where $\alpha$ is the viscosity parameter. Here $c_s = \sqrt{R_g T / \mu}$ is the isothermal sound velocity, where $T$ is the midplane temperature of the disk, $R_g$ is the ideal gas constant, and $\mu = 2.31$ g.mol$^{−1}$ is the mean molar mass of the gas \citep{lo19}. The Keplerian frequency $\Omega_K = \sqrt{G M_\mathrm{\star}/r^3}$ depends on the gravitational constant $G$, the mass of the star $M_\mathrm{\star}$, and the distance from the star $r$. The evolution of the disk is governed by changes in the surface density of the gas $\Sigma_g$ and its midplane temperature $T$. The surface density of the gas is calculated using the equation from \cite{ly74}:

\begin{equation} \label{eq:dSigmaG/dz}
\frac{\partial \Sigma_{\mathrm{g}}}{\partial t} = \frac{3}{r} \frac{\partial }{\partial r} \left [r^{ \frac{1}{2} } \frac{\partial }{\partial r} \left ( r^{ \frac{1}{2} } \Sigma_{\mathrm{g}} \nu \right ) \right ],
\end{equation}

\noindent which we rewrite

\begin{equation}
\begin{split}
\frac{\partial \Sigma_\mathrm{g}}{\partial t} = \frac{1}{2 \pi r} \frac{\partial \dot{M} }{\partial r} \quad \quad \quad \quad \quad \quad \quad  \quad \quad \\
\quad \quad \dot{M} = 3 \pi \Sigma_\mathrm{g} \nu \left ( 1+2 \frac{ \textup{dln}( \Sigma_\mathrm{g} \nu)}{\textup{dln}r} \right ), \quad \quad 
\end{split}
\end{equation}

\noindent where $\dot{M}$ is the mass accretion rate.

We use the following self-similar solutions as initial conditions to solve these equations \citep{ly74,Ag20} :
\begin{equation}
\left \{
\begin{array}{ll}
\Sigma_{\mathrm{g},0} = \frac{\dot{M}_{\mathrm{acc},0}}{3 \pi \nu} e^{-\left (\frac{r}{r_\mathrm{c}} \right )^{0.5} }
\\
\dot{M}_0 = \dot{M}_{\mathrm{acc},0} \left ( 1 - \left ( \frac{r}{r_\mathrm{c}} \right )^{0.5} \right ) e^{-\left (\frac{r}{r_\mathrm{c}} \right )^{0.5} }
\end{array}
\right.
,\end{equation}
\noindent with $r_\mathrm{c}$ the centrifugal radius and $\dot{M}_{\mathrm{acc},0}$ the initial mass accretion rate.

\noindent The midplane temperature is computed by summing all the heating sources \citep{Hu05}: 

\begin{equation} 
\label{eq_appendix:temperature}
\begin{split}
T^4 & = \frac{1}{2 \sigma_\mathrm{sb}} \left ( \frac{3}{8} \tau_\textup{R} + \frac{1}{2 \tau_\textup{P}} \right ) \Sigma_\mathrm{g} \nu \Omega_\mathrm{K}^2  + T_\mathrm{amb}^4
.\end{split}
\end{equation}

\begin{figure}
\includegraphics[width=8cm]{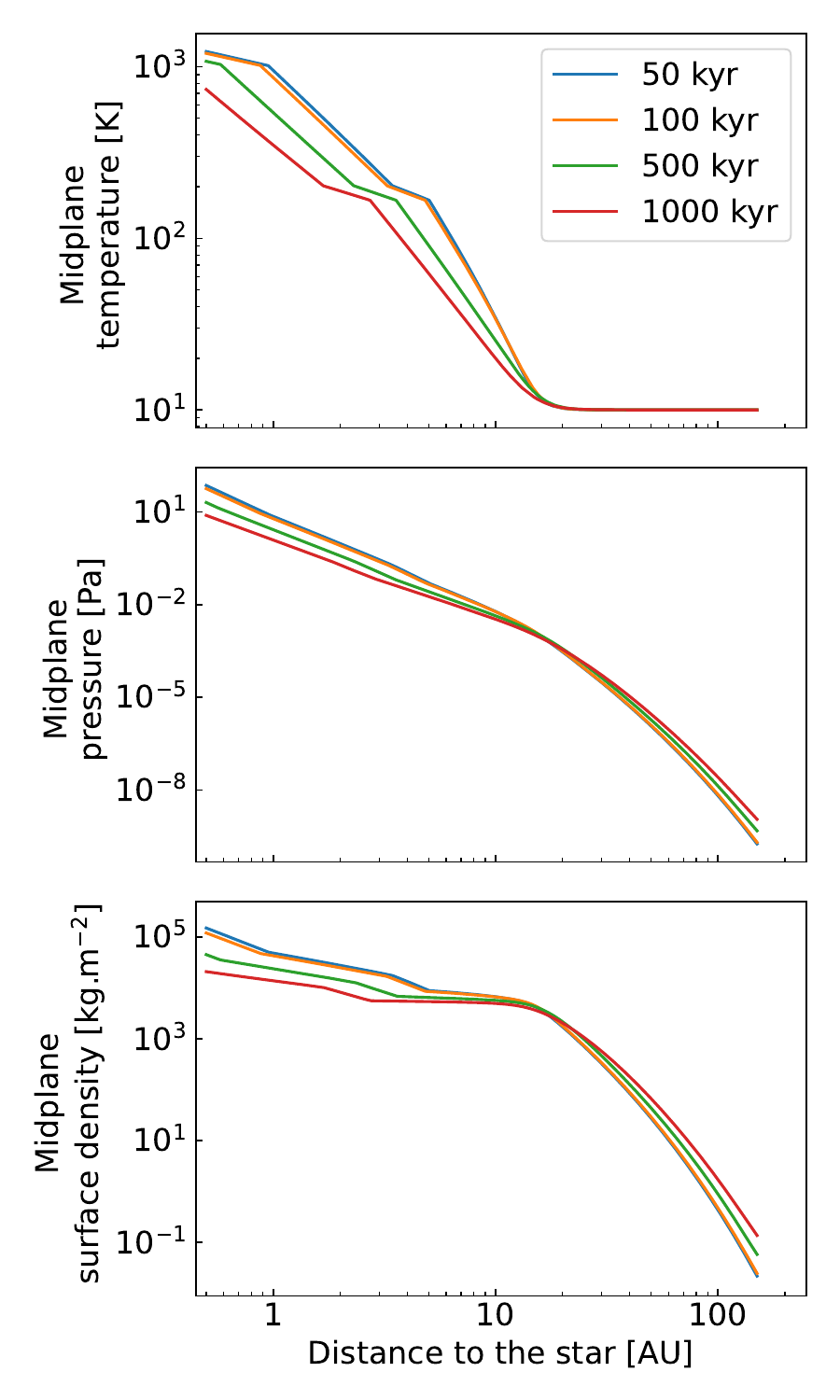}
\caption{Temperature ($T$), pressure ($P$), and surface density ($\Sigma_g$) profiles of the gas in the PSN midplane, shown for various stages of its evolution.}
\label{fig:profiles}
\end{figure}

\noindent The first term represents the viscous heating contribution, where { $\sigma_{sb}$} is the Boltzmann constant, and $\tau_\textup{R}$ and $\tau_\textup{P}$ are the Rosseland and Planck optical depths, respectively. These are related by $\tau _\textup{P} = 2.4 \tau _\textup{R}$ and $\tau _\textup{R} = \frac{\kappa_\textup{R} \Sigma_\mathrm{g}}{2}$, where $\kappa_\textup{R} = \kappa_0 \rho^a T^b$ is the mean Rosseland opacity, with $\rho$ the midplane density and $\kappa_0$, $a$ and $b$ determined from empirical data obtained from Table 3 of \cite{Be94}. { The last term in Eq. \ref{eq_appendix:temperature}} accounts for the contribution of the interstellar ambient temperature $T_\textup{amb} = 10$ K. In the temperature calculation, irradiation from the central star is neglected due to dust accumulation at the iceline positions \citep{Oh21,Ag20,Sc23}. The $\alpha$ parameter is typically set within the range of $10^{-4}$ to $10^{-2}$ \citep{Ha98, Hu05, De17}, while the disk mass accretion rate, $\dot{M}_{\mathrm{acc}}$, is constrained by observations to values between $10^{-9}$ and $10^{-6}\,M_\odot\,\text{yr}^{-1}$ \citep{Ha98, Gu98}. Based on these studies, we define our nominal PSN model using $\alpha = 10^{-3}$ and an initial accretion rate of $\dot{M}_{\mathrm{acc},0} = 5 \times 10^{-7}\,M_\odot\,\text{yr}^{-1}$ \citep{Ha98}. Additionally, we assume an initial disk mass of $0.1 M_\odot$, primarily distributed within a radius of approximately 150 AU \citep{Ag20,Sc23,Mo24}. The midplane temperature, midplane pressure, and surface density profiles of the disk at various evolutionary stages are shown in Fig. \ref{fig:profiles}.

\subsection{Deuterium Chemistry Module}
\label{sec:sec2.2}

The enrichment factor, $f$, is defined as the ratio of the deuterium-to-hydrogen (D/H) ratio in the deuterated species of interest to the D/H ratio in molecular hydrogen. In the case of methane, this ratio can be expressed as:

\begin{equation}
f = \frac{(D/H)_{CH_4}}{(D/H)_{H_2}} = \frac{1}{2}\frac{\frac{CH_3D}{CH_4}}{\frac{HD}{H_2}}.
\label{DH}
\end{equation}

In the following, the D/H ratio in molecular hydrogen is set to the protosolar value, defined as 2.1 $\times$ 10$^{-5}$ \citep{Ha17}. The evolution of $f$ in the protosolar nebula is governed by the following differential equation \citep{Dr99,Mo00,Mo02a}:

\begin{equation}
\frac{\partial f}{\partial r} = k(T) P(A(T) - f) + \frac{1}{\Sigma r} \frac{\partial}{\partial r } (\kappa r \Sigma \frac{\partial f}{\partial r}).
\label{diff}
\end{equation}

\noindent The first term on the right-hand side of Eq.~\ref{diff} represents the isotopic exchange between HD and the deuterated molecule of interest, CH$_3$D. The function $k(T)$ denotes the rate of isotopic exchange, $P$ is the total pressure of the disk, and $A(T)$ represents the equilibrium fractionation factor, which depends solely on temperature $T$. The definition of $f$ is more general than that of $A(T)$, as it does not assume that the system under consideration is in equilibrium. The value of $k(T)$ is obtained from laboratory experiments \citep{Le94,Le96}, while $A(T)$ is derived from quantum mechanical calculations by \cite{Ri77} and subsequently fitted using the analytical expression proposed by \cite{Le96}. Equation~\ref{diff} holds under the assumption that isotopic exchange takes place exclusively between neutral molecules. As shown in \cite{Mo02a}, $k(T)$ decreases by seven orders of magnitude as $T$ decreases from 1000~K to 200~K. This means that at temperatures below 200~K isotope exchange is practically inhibited. At high temperatures, it is also negligible because equilibrium is quickly reached, so $f$ becomes equal to $A(T)$ and the term $(f - A(T))$ converges to 0. The second term on the right-hand side of Eq.~\ref{diff} represents turbulent diffusion within the PSN. This process depends on the local surface density, $\Sigma(R, t)$, and the diffusivity, $\kappa$, defined as the ratio of the turbulent viscosity, $\nu$, to the Prandtl number, $Pr$. The Prandtl number is expressed following the relation provided by \cite{Dr99}:  

\begin{equation}  
Pr = - \frac{4}{3} r \frac{\partial_r \Sigma_g}{\Sigma_g}.  
\label{Pr}  
\end{equation}  

\noindent The value of $Pr$ varies with the opacity of the PSN but remains below unity, which serves as its upper limit \citep{Du91}.

The integration of Eq.~\ref{diff} is governed by the spatial and temporal boundary conditions. For the PSN, the spatial boundary conditions are defined as $\partial f$/$\partial r$~=~0 at $r$~=~1~AU and at the radial distance $r$ corresponding to the condensation or clathration point of methane for any $t$\footnote{$\partial f$/$\partial r$~=~0 at $r$~=~1~AU and the condensation/clathration radius of CH$_4$ implies that the mass flux is neglected at these boundaries due to the poorly constrained physical conditions. This assumption is justified by the rapid decline of the mass flux, which is significant only during the earliest stages of disk evolution. At these early times, high temperatures drive chemical fractionation rather than diffusion, rapidly bringing isotopic enrichment to unity in the inner disk ($T$ $>$ 1000 K) while maintaining its maximum value in the colder outer regions ($T$ $<$ 300 K). At later stages, diffusion becomes the dominant process, but by then, the mass flux is negligible. More details on the mathematical model can be found in \cite{Dr99}.}. The temporal boundary condition is $f(r)$ = \text{constant} at $t$ = 0, applicable to all $r$. This assumption reflects the hypothesis that interstellar ices accreted onto the protoplanetary disk were uniformly enriched throughout the presolar cloud, irrespective of heliocentric distance. Deviations from this assumption would have a negligible impact on the evolution of $f$ across the PSN \citep{Mo02a}. Equation~\ref{diff} applies to the isotopic exchange between CH$_3$D and H$_2$ in the vapor phase. In particular, deuterium exchange is inhibited at temperatures below 200 K. As a result, the evolution of the D/H ratio in the cooling PSN, following the period when the temperature was 200 K, was driven solely by turbulent mixing of the gas. 

The initial value $f_0$, which corresponds to $f$ at $t$ = 0, is poorly constrained. Observations reveal substantial deuterium enrichment in ISM ices \citep{Ce14,Cl16,Pi21}. This enrichment is mainly attributed to significant deuterium fractionation in pre-stellar cores, driven by their extremely low temperatures and the high propensity of molecules to accrete onto icy mantles on dust grains. These processes provide a robust explanation for the elevated deuterium fractionation observed in various molecules within pre-stellar cores \citep{Ro05,Ce14}. Several studies have proposed that interstellar ices vaporized during accretion onto the protoplanetary disk, at heliocentric distances that potentially extend to 30 AU \citep{Ca97}. Upon vaporization, CH$_3$D exchanged deuterium with H$_2$ in the PSN, reducing the deuterium fractionation factor $f$ because the equilibrium values at nebula temperatures are lower than those in the ISM. To date, the only available determination of a purported ``pristine'' D/H ratio in CH$_4$ does not stem from ISM ices, but is instead based on measurements obtained by the ROSINA instrument aboard the Rosetta spacecraft. These measurements, conducted on Comet 67P/C-G, provide a value of \((2.41 \pm 0.29) \times 10^{-3}\) \citep{Mu22}. The central value of this measurement will be used as the default initial condition for our simulations, yielding $f_0$ $\sim$115 relative to the protosolar value.

We consider that CH$_4$ either condensed at approximately 28 K in the PSN or was trapped as clathrate at around 55 K \citep{Mo09a} during the cooling of the PSN. Consequently, the D/H ratio in CH$_4$ within microscopic grains embedded in the nebula, and later in planetesimals, reflects the ratio acquired during the condensation or clathration of methane, depending on the crystallization scenario considered. In the following, our calculations consistently represent the D-enrichment value acquired by methane at the point of condensation or clathration in the PSN.


\section{Results} 
\label{sec:sec3}

\begin{figure*}
\centering
\includegraphics[width=18cm]{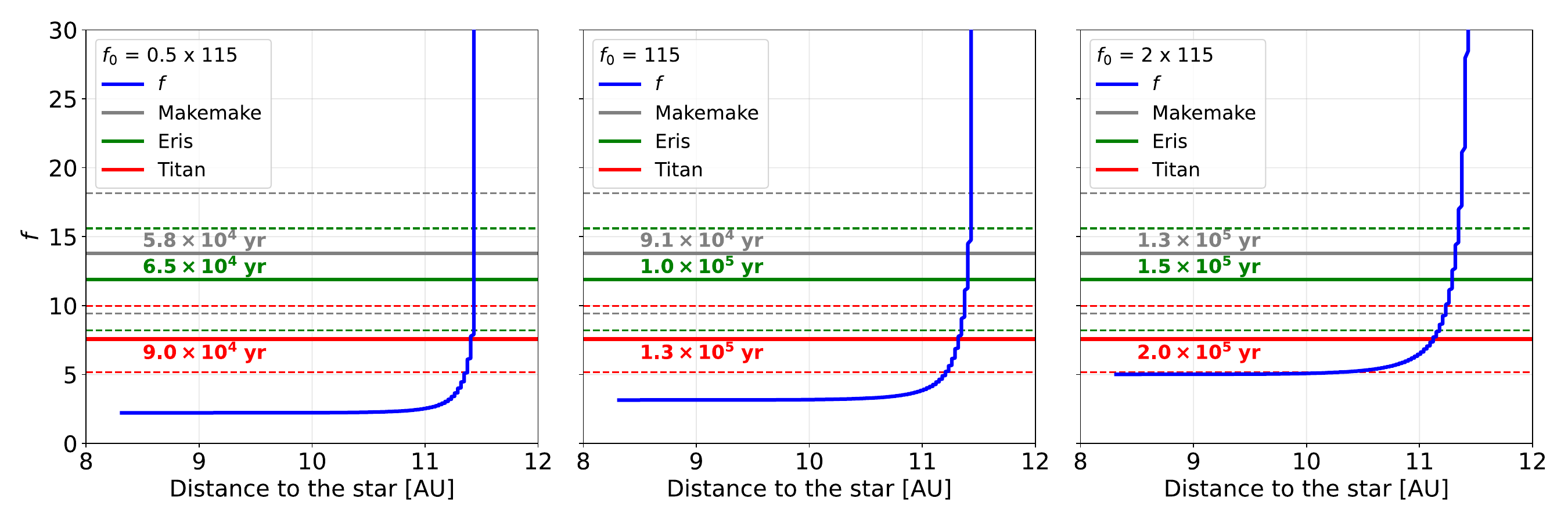}
\caption{Deuterium enrichment profile $f$ calculated along the trajectory of the methane iceline in the PSN, assuming a condensation temperature of 28 K. The thick horizontal lines indicate the nominal enrichment values observed for Makemake (gray), Eris (green), and Titan (red). Dashed lines in matching colors represent the one--$\sigma$ uncertainties associated with these measurements. From left to right, the initial enrichment values were set at 0.5, 1, and 2 times the central value of 115 measured in Comet 67P/C-G. The epochs where $f$ intersects the central values observed for the three bodies are highlighted, corresponding to a very narrow variation in the location of the methane iceline within the PSN.}
\label{fig28K}
\end{figure*}

\begin{figure*}
\centering
\includegraphics[width=18cm]{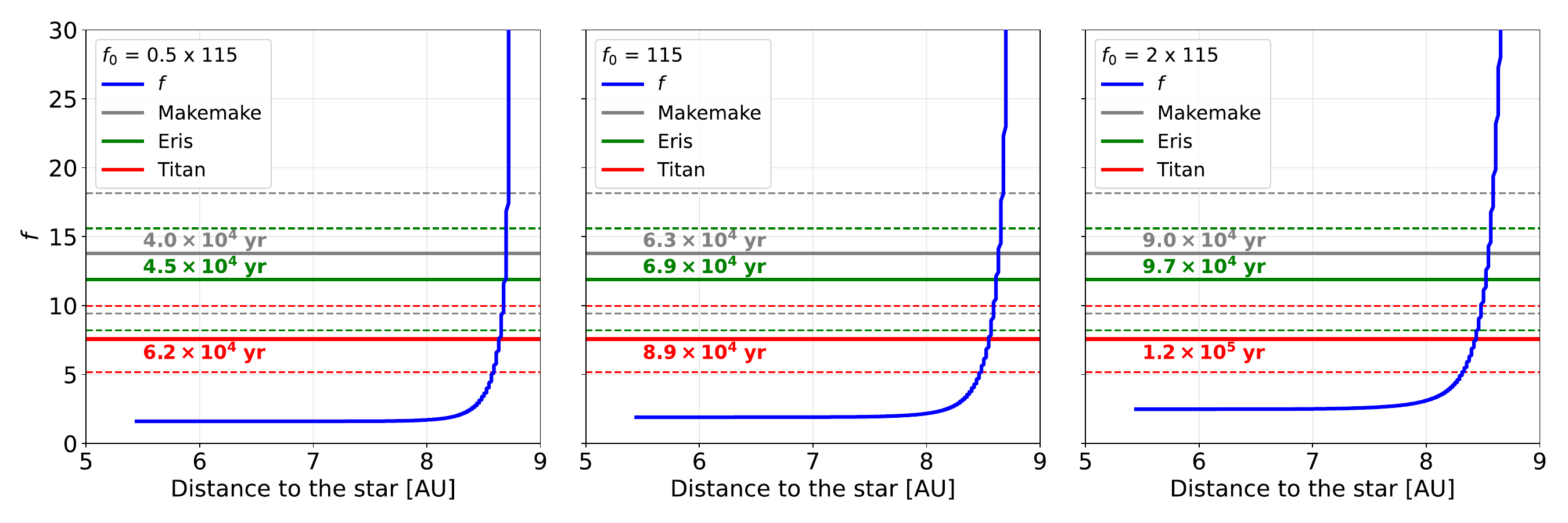}
\caption{Similar to Fig. \ref{fig28K}, but showing the deuterium enrichment profile $f$ calculated along the trajectory of methane clathration in the PSN, with an assumed entrapment temperature of 55 K.}
\label{fig55K}
\end{figure*}

Figures \ref{fig28K} and \ref{fig55K} show the evolution of the $f$--profile for CH$_4$ in the PSN, as determined from our nominal model. The profiles are calculated along the trajectories of the methane iceline and methane clathrate line, corresponding to condensation and entrapment temperatures of 28 K and 55 K, respectively. These calculations are compared with D/H measurements in CH$_4$ obtained by the Cassini spacecraft in the Titan atmosphere and by the JWST on the surfaces of the KBOs Makemake and Eris (see Table \ref{tab:data}). We consider three initial conditions, corresponding to 0.5, 1, and 2 times the central enrichment measured on Comet 67P/C-G. The deuterium enrichment profiles show a sharp gradient between the 11--12 AU and 8--9 AU regions, before leveling off in the inner regions of the PSN. This behavior is attributed to the slow temporal evolution of the disk's thermodynamic profiles, which causes a gradual inward progression of the two considered icelines. The slow progression allows sufficient time for efficient isotopic exchange to take place in the gas phase before methane either condenses or becomes trapped in clathrates.


The calculated PSN enrichment profiles agree consistently with all observed measurements. The central values measured in Makemake, Eris, and Titan intersect the $f$--profile at different time intervals. For Makemake, the intersections occur at 5.8 $\times$ 10$^4$, 9.1 $\times$ 10$^4$, and 1.3 $\times$ 10$^5$ yr. For Eris, they occur at 6.5 $\times$ 10$^4$, 1.0 $\times$ 10$^5$, and 1.5 $\times$ 10$^5$ years. For Titan, the intersections occur at 9.0 $\times$ 10$^4$, 1.3 $\times$ 10$^5$, and 2.0 $\times$ 10$^5$ yr. These values { at the methane iceline} correspond to $f_0$ values of 0.5, 1, and 2 times the central value of 115 measured in Comet 67P/C-G. Our calculations assume that the three bodies accreted icy solids crystallized in a narrow region of the PSN, typically spanning a few tenths of an AU. This region lies between 11--12 AU and 8--9 AU, depending on whether solid methane forms from pure condensate or clathrate. Our results support the hypothesis that the building blocks of giant planet systems and KBOs originated within a compact region of the PSN, in agreement with predictions from dynamical models \citep{Ts05,Ne18}. This interpretation aligns with the D/H ratio measured in H$_2$O on Enceladus, a moon of Saturn located closer to the planet than Titan. Data from the Cassini-Huygens mission revealed a cometary D/H ratio ($\sim$2.9 $\times$ 10$^{-4}$), suggesting that the building blocks of Enceladus formed in a region of the PSN similar to that of many comets \citep{Wa09,Mo09c,Clar19}. This conclusion also extends to Titan, as it accreted in a colder region of Saturn's circumplanetary disk from building blocks that formed earlier in the PSN \citep{Mou02}. 

 Additional tests assuming up to five times the central enrichment measured in Comet 67P/C-G were also performed and yielded consistent results. This suggests that our model effectively reproduces the D/H ratios observed in Titan's atmosphere and on the surfaces of Eris and Makemake, without requiring an internal source of CH$_4$ from hydrothermal or metamorphic processes \citep{Mo09a,Mo09b,At06,Gl24}.

\begin{table}[htpb]
\centering
\caption{D/H measurements in CH$_4$ in 67P/C-G, Titan, Eris, and Makemake}
\begin{tabular}{lcc}
\hline
\hline
\smallskip
Object              & Value                                 & Reference\\
\hline
67P/C-G             & (2.41 $\pm$ 0.29) $\times$ 10$^{-3}$    & \cite{Mu22}\\
Titan               & (1.59 $\pm$ 0.33) $\times$ 10$^{-4}$  & \cite{Ni12} \\
Eris                & (2.5 $\pm$ 0.5) $\times$ 10$^{-4}$    & \cite{Gr24} \\
Makemake            & (2.9 $\pm$ 0.6) $\times$ 10$^{-4}$    & \cite{Gr24} \\
\hline
\end{tabular}
\label{tab:data}
\end{table}

\section{Discussion and Conclusions} 
\label{sec:sec4}

The deuterium enrichments in CH$_4$ observed on Titan, Makemake, and Eris can be matched by our model for any value of the initial accretion rate, $\dot{M}_{\mathrm{acc},0}$, in the range $5 \times 10^{-7}$ to $10^{-6}\,M_\odot\,\text{yr}^{-1}$, assuming { $\alpha$ = $10^{-3}$}. At constant $\alpha$, higher accretion rates result in a warmer protoplanetary disk, causing the deuterium enrichment profile to shift outward to higher heliocentric distances while maintaining a similar overall shape. For example, assuming $\alpha$ = $10^{-3}$ and an initial accretion rate of 10$^{-6} M_\odot\,\text{yr}^{-1}$ instead of $5 \times 10^{-7}\ M_\odot\,\text{yr}^{-1}$, the $f$--value crossings for the central measurements of Titan, Makemake, and Eris would shift about 2~AU farther from the Sun. On the other hand, an initial accretion rate lower than $5 \times 10^{-7}$ would result in colder temperatures in the outer regions of the PSN, preventing the $f$--profile from decreasing to the values observed in the three bodies.

The calculated $f$--profiles also show significant sensitivity to variations in the $\alpha$--parameter. Our results indicate that for any $\alpha$--value greater than $10^{-3}$, the $f$--profile does not decrease sufficiently to simultaneously intersect all three measurements. Specifically, with $\alpha = 2 \times 10^{-3}$ and the same initial accretion rate as in our nominal model, the $f$--profile only reaches the upper limit of Makemake's D/H measurement. This is because the $f$--value depends on turbulent viscosity (see Eq. \ref{diff}), and higher values of $\alpha$ lead to a higher enrichment of deuterium for the same set of other parameters.

Our model allows for some tuning through the viscosity and accretion rate parameters, enabling a range of D/H ratios between the protosolar value and that measured in 67P. However, this flexibility is intrinsic to disk evolution models, where temporal evolution naturally governs isotopic fractionation. The model remains constrained by physical processes such as vapor diffusion, isotopic exchange, and condensation, which shape the range of possible outcomes.

A key test of our approach would be independent measurements of D/H ratios in methane across different cometary bodies. If comets were found to exhibit D/H ratios in methane comparable to those measured on Makemake and Eris, or intermediate between the protosolar value and that observed in 67P, this would provide a strong consistency check of the D/H pattern predicted by our model and a foundation for further observational validation. Additionally, a precise determination of the D/H ratio in interstellar methane would be { helpful} for refining the initial conditions of our simulations.

Our model also predicts the formation of CH$_4$--rich ices after approximately 60–150 kyr of PSN evolution, with deuterium enrichments matching the values observed in Makemake and Eris. As the CH$_4$ iceline moves inward, CH$_4$--rich microscopic grains form, agglomerate with other grains, and grow into pebbles. Once the abundance of icy pebbles exceeds a critical threshold near the iceline, planetesimal formation is triggered via streaming instability, with timescales reaching several Myr for planetesimals of the order of 100 km in size \citep{Jo15,Dr17,Le20}.

The high D/H ratio measured in CH$_4$ in 67P/C-G aligns with the previously reported D/H value in H$_2$O ((5.3 ± 0.7) $\times$ 10$^{-4}$) by \cite{Al15}, suggesting that the comet was assembled from nearly pristine materials that experienced minimal reprocessing in the PSN. However, a recent analysis by \cite{Ma24} revises the D/H ratio in H$_2$O downward to $\sim$(2.2–2.9) $\times$ 10$^{-4}$ --namely 1.2–1.6 times the terrestrial value-- after accounting for dust contamination that may have artificially elevated the original measurement from the ROSINA instrument. If confirmed, this revision implies that the comet’s water underwent significant isotopic exchange with hydrogen before crystallizing in the PSN, whereas CH$_4$ remained linked to a more primitive reservoir. This coexistence could result from large-scale mixing of solid materials across the PSN, a process already invoked to explain isotopic heterogeneity in meteorites { \citep{Del97,Del98}.}

An important marker of primordial ice is CO, which is not detected on Eris and Makemake according to \cite{Gr24}'s analysis of the JWST data. To reconcile a primordial origin of CH$_4$ with an { apparent} absence of CO, several chemical and physical processes must be considered. One possibility, as suggested by \cite{Gl18}, is that CO was destroyed by reactions in a subsurface liquid-water ocean. 
Another possibility is that CH$_4$ was preferentially enriched at its iceline in the PSN relative to CO and other volatiles, contributing to the observed depletion \citep{Mou21}. This hypothesis supports the near absence of detected volatiles other than CH$_4$ and N$_2$ on Makemake and Eris \citep{Gr24}, with the observed nitrogen possibly coming from organic matter { \citep{Mil19}.}

Our model suggests that the methane observed on Eris and Makemake could originate from the PSN while not excluding the possibility of endogenous production within these bodies \citep{Gl24}. Although this does not confirm endogenous methane production, the differentiation of Eris \citep{Ni23} and the potential cryovolcanic activity on Makemake \citep{Ki24} are in agreement with this hypothesis. However, our study indicates the potential coexistence of two methane reservoirs: one primordial and another produced in situ. Furthermore, our findings support the idea that these KBOs accreted from solids formed in an environment with a high diversity of ices.

Finally, our model focuses only on gas diffusion and does not consider the transport of grains and pebbles with different D/H ratios across the disk. This contrasts with the detection of crystalline silicates in comets, suggesting their formation near the protosun followed by efficient radial transport \citep{Bo02,Mou07,Br14}. Addressing this limitation is an important direction for future work, as the resulting deuterium distribution in solids formed from dust and pebbles may differ from the idealized patterns presented here. A similar consideration applies to \cite{Gl24}, who proposed that the D/H ratio in H$_2$O and CH$_4$ remains constant in bodies that accreted these volatiles from the PSN. This assumption { seems questionable}, as it assumes identical chemical pathways for both molecules from the ISM to the PSN and neglects the potential effects of radial mixing on their pure condensates.\\

The project leading to this publication has received funding from the Excellence Initiative of Aix-Marseille Universit\'e--A*Midex, a French ``Investissements d’Avenir program'' AMX-21-IET-018. This research holds as part of the project FACOM (ANR-22-CE49-0005-01\_ACT) and has benefited from a funding provided by l'Agence Nationale de la Recherche (ANR) under the Generic Call for Proposals 2022. OM acknowledges funding from CNES.

\bibliography{deut}



\end{document}